\documentclass{article}
\usepackage{spconf,amsmath,graphicx,hyperref}
\usepackage{url}  
\usepackage{multirow, booktabs, bbding, enumerate, enumitem, subcaption, xcolor} 
\usepackage{amsmath, amssymb}
\usepackage{graphicx}
\usepackage{tablefootnote}

\title{MMAudioSep: Taming Video-to-Audio Generative Model towards Video/Text-Queried Sound Separation}
\name{
    Akira Takahashi$^{1}$,\thanks{\dag~We would like to thank Marc Ferras and Koo Junghyun for their valuable comments during the preparation of this manuscript. We also appreciate the helpful discussions with Keisuke Toyama and Zhi Zhong.}
    Shusuke Takahashi$^{1}$,
    Yuki Mitsufuji$^{1,2}$}
\address{
$^{1}$Sony Group Corporation, Japan \;
$^{2}$Sony AI, USA\
}
\begin{document}
\ninept 
\maketitle
\begin{abstract}

We introduce MMAudioSep, a generative model for video/text-queried sound separation that is founded on a pretrained video-to-audio model. By leveraging knowledge about the relationship between video/text and audio learned through a pretrained audio generative model, we can train the model more efficiently, i.e., the model does not need to be trained from scratch. We evaluate the performance of MMAudioSep by comparing it to existing separation models, including models based on both deterministic and generative approaches, and find it is superior to the baseline models. Furthermore, we demonstrate that even after acquiring functionality for sound separation via fine-tuning, the model retains the ability for original video-to-audio generation. This highlights the potential of foundational sound generation models to be adopted for sound-related downstream tasks. Our code is available at \url{https://github.com/sony/mmaudiosep}.
\end{abstract}
\vspace{-2mm}
\section{Introduction}
\label{sec:intro}
\vspace{-1mm}
Sound separation (SS) models that use conditional information to control model behavior have been widely discussed, including visual-queried separation \cite{dong2023clipsep, Chen2023iQuery}, text-queried separation \cite{liu22w_interspeech,liu2023separate,yuan2024flowsep}, and omni-modality separation \cite{cheng2025omnisep}. Recently, sound separation based on a generative approach \cite{yuan2024flowsep} has been presented, whereas most sound separation models have been implemented as discriminative tasks.

Recently, a model for audio synthesis based on video and text queries has demonstrated plausible results and is able to generate audio signals that correspond semantically and temporally to the input video and text. Among them, MMAudio \cite{cheng2025taming} synthesizes high-quality and synchronized audio given video and optional text conditions using a novel multimodal joint training framework. 
However, these two research directions, generative sound separation and video-to-audio generation, have developed independently. While generation models possess rich multimodal knowledge that could potentially benefit separation tasks, this cross-domain knowledge transfer remains unexplored. 
We hypothesize that the knowledge stored in the synthesis model can be applied to the sound separation model.

\begin{figure}[t]
\begin{minipage}[b]{1.0\linewidth}
  \centering
  \centerline{\includegraphics[width=6.cm]{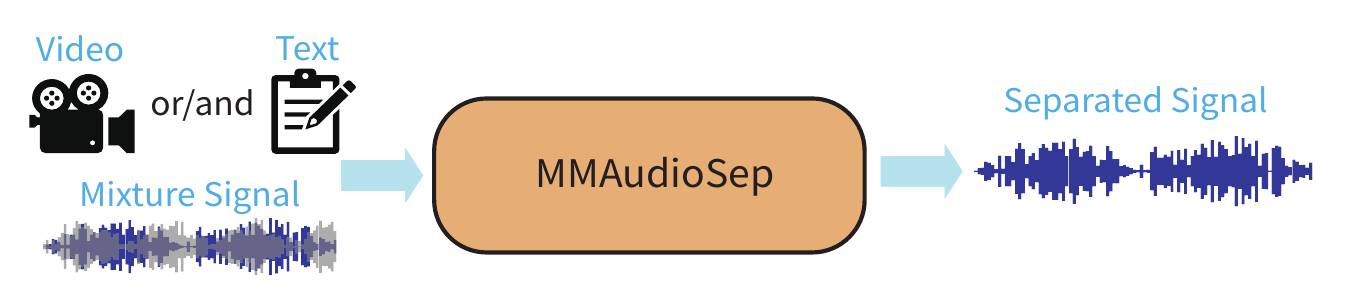}}
  \vspace{-1mm}
  \centerline{(a) Sound Separation}\medskip
\vspace{-2mm}
\end{minipage}
\vspace{-2mm}
\begin{minipage}[b]{1.0\linewidth}
  \centering
  \centerline{\includegraphics[width=6.cm]{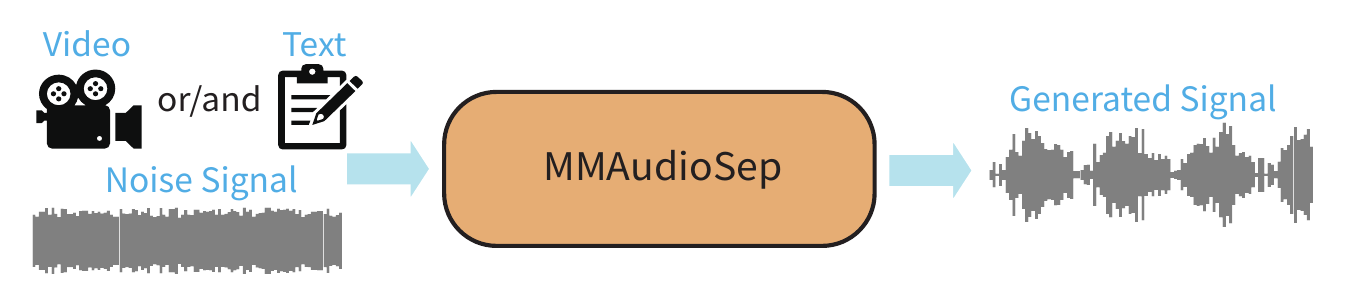}}
  \vspace{-1mm}
  \centerline{(b) Video-to-Audio Generation}\medskip
\end{minipage}
\hfill
\vspace{-4mm}
\caption{The concept of this paper. Both (a) and (b) utilize the MMAudioSep model with identical parameters.}\label{fig:concept}
\vspace{-4mm}
\end{figure}
In this study, we propose a fine-tuning method for a visual/text-queried sound separation model based on a pretrained video-to-audio generation model, expecting that the adapted model will exploit the knowledge of the pretrained video-to-audio model.
We applied fine-tuning to the pretrained MMAudio model to construct a visual/text-queried sound separation model on the basis of a generative approach. The pretrained MMAudio takes video and text as inputs and automatically generates sound that corresponds to the input queries, using a flow-matching algorithm. We incorporated additional sound input, which contained a sound mixture, into the input of the MMAudio architecture and considered the network output as the separated result. We then applied fine-tuning with the training data for sound separation. 
To the best of our knowledge, this is the first research that enables video-to-audio generation and query-based sound separation in a single model.
Our threefold contributions are summarized as follows: 

\begin{itemize}[leftmargin=18pt]
\item We propose a fine-tuning approach that adapts pretrained video-to-audio models for video/text-queried sound separation.
\item We demonstrate that MMAudioSep outperforms existing video/text-queried sound separation models.
\item We show that even after fine-tuning for sound separation, MMAudioSep retains the ability for video-to-sound synthesis.
\end{itemize}
Fig. \ref{fig:concept} illustrates the conceptual framework of our approach.
\vspace{-2mm}
\section{Related Work}
\label{sec:related}
\vspace{-1mm}
\subsection{Sound Separation}\label{ssec:overview}
\vspace{-1mm}
Sound separation reconstructs individual sound sources from mixed signals, crucial for multi-speaker environments, music production, and assistive listening \cite{araki2025source}.  Recent advances include source extraction using auxiliary information and hybrid methods that combine signal processing with neural networks. \\
\textbf{Text-Queried Sound Separation (TQSS)} enables separation through natural language prompts. LASS-Net \cite{liu22w_interspeech} aligns audio and text embeddings for separation. AudioSep \cite{liu2023separate}, trained on over 14,000 hours, improves separation across sound types. We propose a multimodal framework using visual and textual inputs for enhanced separation. \\
\textbf{Image-Queried Sound Separation (IQSS)} uses visual guidance for separation. Sound of Pixels \cite{Zhao_2018_ECCV} linked object appearance to sound components. Later works \cite{gao2019cosep,Zhao_2019_ICCV,zhu2021amnet} added motion cues and self-supervised learning. iQuery \cite{Chen2023iQuery} reformulated IQSS as query-based segmentation. CLIPSep \cite{dong2023clipsep} uses CLIP \cite{radford2021clip} text embeddings for zero-shot separation. \\ 
\textbf{Generative Approach for Sound Separation.} FlowSep \cite{yuan2024flowsep} is a generative model for language-queried separation using Rectified Flow Matching. FlowSep learns linear flow trajectories from Gaussian noise to target source features, which are decoded into mel spectrograms. Trained on 1,680 h of audio, FlowSep achieved state-of-the-art performance. Our method generates source signals from latent representations with multimodal inputs. \\
\textbf{Multimodal Sound Separation.} OmniSep \cite{cheng2025omnisep} extracts target sounds using audio, image, and text queries. It employs Query-Mixup for cross-modal representations and supports both positive and negative queries. OmniSep achieves state-of-the-art performance across query modalities. Like OmniSep, our approach integrates multimodal information for sound separation. 
\vspace{-2mm}
\section{MMAudioSep}
\label{sec:proposal}
\vspace{-1mm}
\subsection{MMAudio}
\label{ssec:mmaudio}
\vspace{-1mm}
\begin{figure}[t]
  \includegraphics[width=\columnwidth]{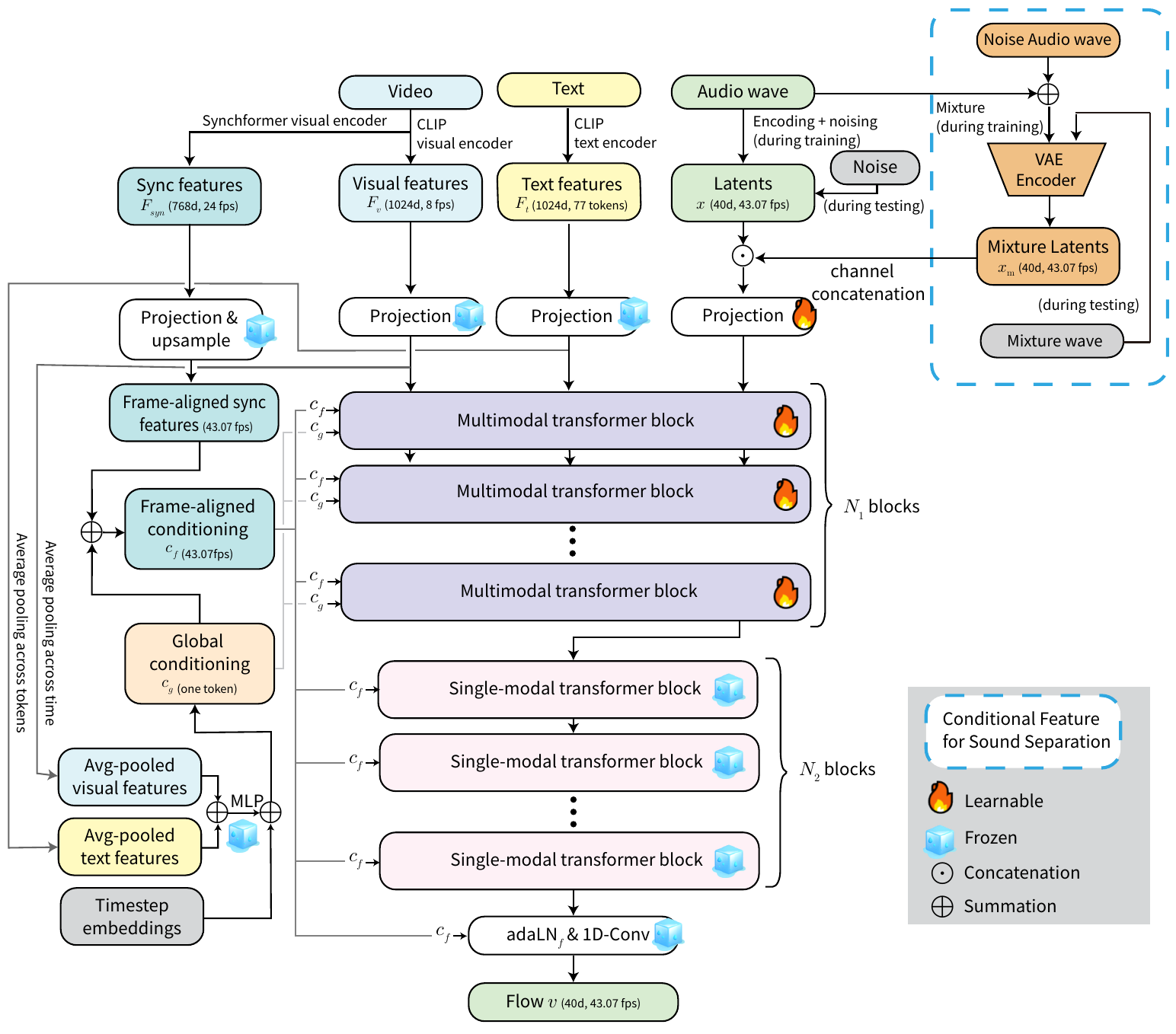}
  \vspace{-4mm}
  \caption{The Overview of MMAudioSep Architecture}\label{fig:arch}
  \vspace{-6mm}
\end{figure}
MMAudio \cite{cheng2025taming} is a generative model that uses flow matching to synthesize audio of an input video with optional text conditions. The model operates in a latent space where audio waveforms are encoded via a pretrained variational autoencoder (VAE). In this section, we describe the overview of MMAudio, which serves as the foundation for our proposed MMAudioSep.

Fig. \ref{fig:arch} includes the architecture of MMAudio, which is based on a multimodal transformer network (MM-DiT) \cite{esser2024scaling} with input of video, text, and audio. 
The video input consists of semantic features from CLIP (8 fps, 1024-dim) and audio-visual synchronization features from Synchformer (24 fps, 768-dim).
The text input uses CLIP features (77 tokens, 1024-dim). Audio is represented as latent vectors from a VAE (43 fps, 40-dim).
In MMAudio, input audio information is used only during training, while Gaussian noise serves as the starting point during inference. Visual features and text features are converted through linear projection layers into hidden dimension $h$. Audio features are also converted into hidden dimension $h$. These hidden representations are passed through $N_1$ layers of multimodal transformer blocks, followed by audio-only transformer blocks with $N_2$ layers, which generate audio latent vectors. To achieve further controllability with video and text, video and text features are pooled and injected into the network through adaptive layer normalization (adaLN) layers as global conditioning. Additionally, frame-aligned conditioning from Synchformer features is used to enhance audio-visual synchronization.
During testing, the generated latents are decoded by the VAE into spectrograms that are then vocoded by the pretrained BiGVGAN vocoder \cite{sang2023bigvgan} into audio waveforms.

MMAudio uses the conditional flow matching (CFM) objective for training. 
At test time, to generate a sample, a noise sample $x_0$ is randomly drawn from the standard normal distribution, and an ordinary differential equation (ODE) solver is used to numerically integrate from time $t=0$ to time $t=1$ following a learned time-dependent conditional velocity vector field $v_\theta (t, C, x) : [0, 1] \times \mathbb{R}^C \times \mathbb{R}^d \rightarrow \mathbb{R}^d$, where $t$ is the timestep, $C$ is the condition (e.g., video and text), and $x$ is a point in the vector field. The velocity vector field is represented via a deep net parameterized by $\theta$.
During the training phase, the parameter $\theta$ is optimized by evaluating the CFM objective, expressed as \begin{equation} L_\text{CFM}(\theta)=\mathbb{E}_{t, q(x_0), q(x_1, C)} \lVert v_\theta (t, C, x_t) - u(x_t|x_0, x_1) \rVert ^2, \label{eq:cfm_objective}\end{equation} where $t\in [0, 1]$, $q(x_0)$ represents the standard normal distribution, and $q(x_1, C)$ is derived from the training dataset. Furthermore, the equation $x_t = t x_1 + (1 - t) x_0$ delineates a linear interpolation trajectory between noise and data. Additionally, $u(x_t | x_0, x_1) = x_1 - x_0$ specifies the corresponding flow velocity at $x_t$.
\begin{figure}[t]
  \centering
  \includegraphics[width=0.85\columnwidth]{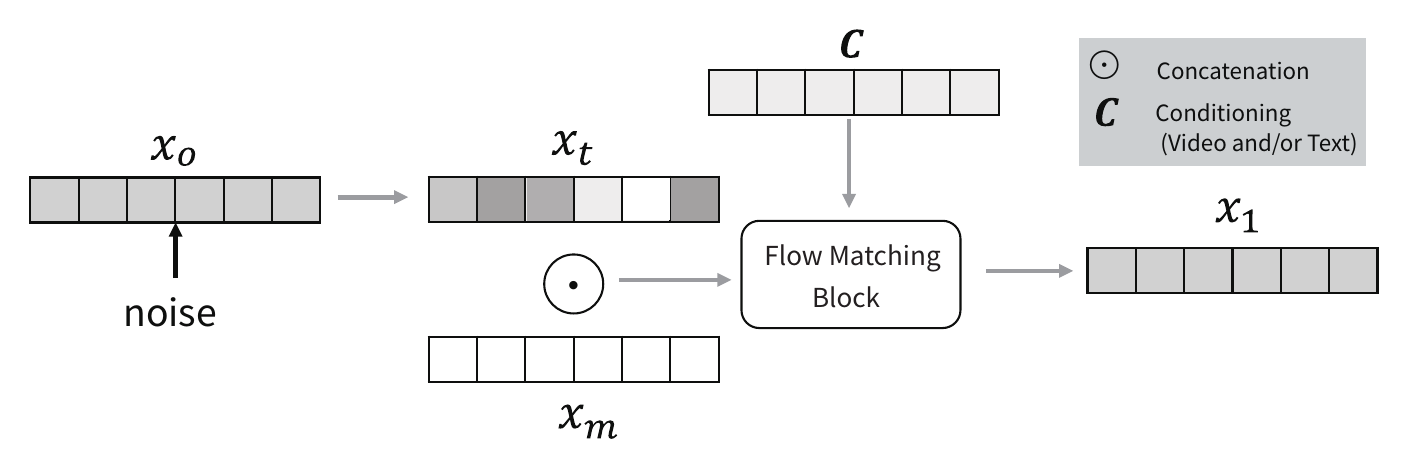}
  \vspace{-4mm}
  \caption{Channel-concatenation conditioning mechanism}
  \label{fig:ccm}
  \vspace{-2mm}
\end{figure}
\vspace{-2mm}
\subsection{MMAudioSep}
\label{ssec:mmaudiosep}
%
Fig. \ref{fig:arch} shows the network architecture of MMAudioSep, which closely follows that of MMAudio, except for the modified audio projection layer. This change alters the input channel information of the projection layer through a channel-concatenation conditioning mechanism.
During fine-tuning, we update parameters for the audio projection layer, multimodal transformer blocks, while other parameters remain frozen.\\
\textbf{Channel-concatenation conditioning mechanism}.
\label{ssec:conditioning}
MMAudioSep aims to generate the target feature $\hat{x}_1$, which represents the separated sound, given the mixture waveform along with visual and text queries. Inspired by FlowSep \cite{yuan2024flowsep}, we apply channel-conditioned generation to guide the model using the mixture as input channel conditions. As shown in Fig. \ref{fig:ccm}, both noise from standard Gaussian distribution $x_0$ and the latent vectors of mixture mel-spectrogram $x_m$ are concatenated along the channel dimension before being forwarded into the MMAudio module. In this way, the additional condition, $x_m$, is considered as extra information within the input so that both the mixture latent vector and the target latent vector can be processed in parallel by the flow matching model. The noise adding forward process only affects the target latent vector, while the mixture channel $x_m$ remains unchanged. At training time, we use CFM objective Eq. \eqref{eq:cfm_objective}, which is the same as that of MMAudio. \\
\textbf{Model Architecture.}
We developed MMAudioSep utilizing the 44k-large variant of MMAudio, selected for its superior capability to capture audio with high-resolution fidelity. This variant functions at a 44.1 kHz sampling rate, enabling the model to retain the fine-grained acoustic details that are frequently lost in lower-resolution formats. Building upon this high-fidelity foundation, we trained our model to execute video/text-queried sound separation. 
\begin{figure}[t]
  \centering
  \includegraphics[width=0.9\columnwidth]{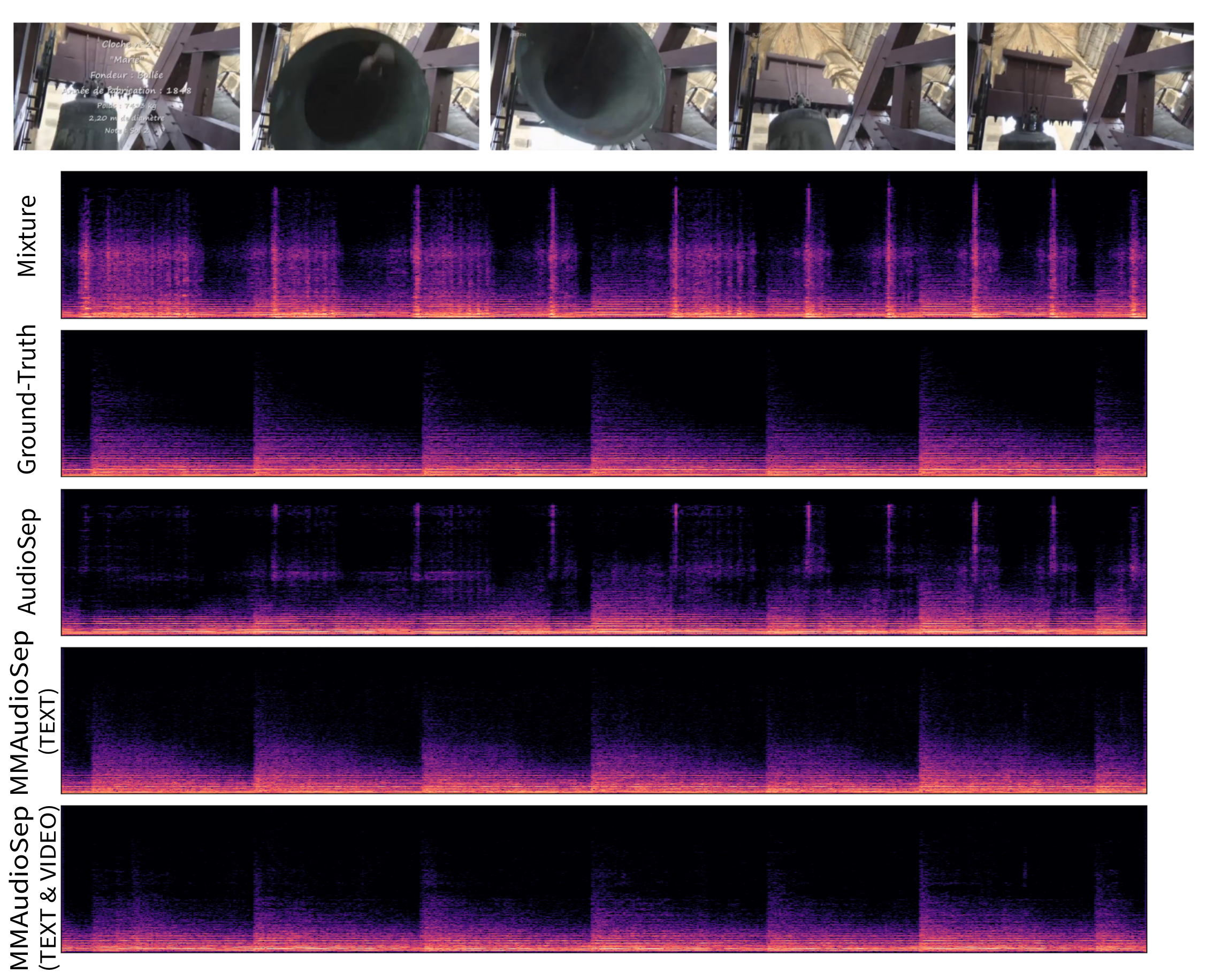}
  \vspace{-2mm}
  \caption{Visualize the spectrograms of the mixture, ground-truth, and separated audio (AudioSep and Our model). "church bell ringing."}\label{fig:ex_ss_result}
  \vspace{-2mm}
\end{figure}
\begin{table*}[ht]
\centering
\caption{Source Separation Benchmarking on VGGSound-Clean and MUSIC test set. The best scores are bolded. CLAP and CLAP-A refer to CLAPScore and CLAPScore$_\text{A}$, respectively. }
\label{tab:res_main} 
\vspace{-2mm}
\resizebox{0.93\linewidth}{!}{
\begin{tabular}{l|cc|cccccc|cccccc}
\hline
 & \multicolumn{2}{c|}{Query} & \multicolumn{6}{c|}{VGGSound-Clean}& \multicolumn{6}{c}{MUSIC} \\
 \cmidrule(lr){4-9} \cmidrule(lr){10-15}
Method &  TEXT & VIDEO &  FAD$\downarrow$ & IS$\uparrow$ & CLAP $\uparrow$ & CLAP-A $\uparrow$ & IB-Score $\uparrow$ & DeSync $\downarrow$ & FAD $\downarrow$ & IS $\uparrow$ & CLAP $\uparrow$ & CLAP-A $\uparrow$ & IB-Score $\uparrow$ & DeSync $\downarrow$ \\
\midrule
Ground Truth&-&-&-&15.84&29.13&100.0&36.15&0.665& - &3.40&32.81&100.0&44.45& 0.176 \\
\midrule
Mixture&-&-&2.51&7.53&17.05&59.34&18.85&0.906&3.70&2.48&26.90&75.18&33.75& 0.345 \\
\midrule
AudioSep& \checkmark & - &\textbf{0.90} &12.82&28.37&79.79&29.94&0.787&1.37&3.08&31.11&86.69&40.74&0.254  \\
\midrule
FlowSep & \checkmark & - &1.90&10.01&24.79&56.99&25.22&0.840&18.87&3.17&24.33&49.02&19.69&0.648 \\
\midrule
MMAudioSep & \checkmark & - & 1.31 & 13.41 & 29.37 & 77.24 & 29.80 & 0.813 & 1.14 & 3.31 & 29.50 & 85.90 & 39.26 & 0.345 \\
(scratch) & \checkmark & \checkmark & 1.88 & 13.05 & 29.30 & 80.69 & 33.18 & 0.667 & \textbf{0.97} & 3.36 & 32.10 & 91.11 & 43.08 & 0.212 \\
\midrule
MMAudioSep & \checkmark & - &  1.15 & \textbf{14.23} & 30.57 & \textbf{81.46} & 31.38 & 0.789 & 1.12 & 3.28 & 30.80 & 88.28 & 40.63 & 0.289 \\
(pretrain wo/frozen) & \checkmark & \checkmark & 1.83 & 13.69 & \textbf{30.59} & 81.35 & 34.06 & 0.661 & 1.18 & \textbf{3.43} & \textbf{32.48} & \textbf{91.48} & \textbf{44.21} & 0.182 \\
\midrule
MMAudioSep & \checkmark & - & 1.36 & 13.56 & 30.02 & 79.43 & 30.41 & 0.823 & 1.22 & 3.29 & 30.34 & 87.44 & 40.64 & 0.287 \\
(pretrain w/frozen)& \checkmark & \checkmark & 1.98 & 13.87 & 30.38 & 80.11 & \textbf{34.23} & \textbf{0.629} & 1.72 & 3.42 & 31.69 & 90.29 & 44.06 & \textbf{0.179} \\
 \hline
\end{tabular}
}
\end{table*}
\vspace{-1mm}
\begin{table*}[ht]
\centering
\caption{V2A Generation Benchmarking on VGGSound test set. The parameter counts exclude pretrained feature extractors (e.g., CLIP), latent space encoders/decoders, vocoders, and frozen parameters. The top-3 scores are bolded.}
\label{tab:res_vgg_all}
\vspace{-2mm}
\resizebox{0.62\linewidth}{!}{
\begin{tabular}{lc|cccc|cc}
\toprule
Method&Params& FD $\downarrow$ & FAD $\downarrow$ & KL $\downarrow$ & IS $\uparrow$ & IB-Score $\uparrow$ & DeSync $\downarrow$ \\
\midrule
ReWaS \cite{jeong2024rewas}        &  619M & 141.38  & 1.79 & 2.87 &  8.51 &  14.82 & 1.062 \\
Seeing\&Hearing \cite{xing2024seeingandhearing} & 415M & 219.01 & 5.40 & 2.26 &  8.58 &  \textbf{33.99} & 1.204 \\
V-AURA \cite{viertola2025vaura} & 695M & 218.50 & 2.88 & 2.42 & 10.08 & 27.64 & 0.654 \\
VATT \cite{liu2024vatt} & - & 131.88 & 2.77 & \textbf{1.48} & 11.90 & 25.00 & 1.195 \\
Frieren \cite{wang2024frieren} & \textbf{159M} & 106.10 & 1.34 & 2.73 & 12.25 & 22.78 & 0.851 \\
FoleyCrafter \cite{zhang2024foleycrafter} & 1.22B & 140.09 & 2.51 & 2.30 & \textbf{15.68} & 25.68 & 1.225 \\
V2A-Mapper \cite{wang2024v2amapper} & \textbf{229M} & \textbf{84.57} & \textbf{0.84} & 2.69 & 12.47 & 22.58 & 1.225 \\
SpecMaskFoley \cite{zhong2025specmaskfoley} & \textbf{300M} & 108.62 & \textbf{1.03} & 1.76 & 11.43 & 26.39 & 0.652 \\ 
MMAudio-L-44k \cite{cheng2025taming} & 1.03B & \textbf{60.60} & \textbf{0.97} & \textbf{1.65} & \textbf{17.40} & \textbf{33.22} & \textbf{0.442} \\
\midrule
MMAudioSep (scratch)  & 1.03B & 146.05 & 2.42 & 2.11 & 10.31 & 25.09 & 0.655 \\
MMAudioSep (pretrain wo/frozen) & 1.03B & 104.84 & 1.68 & 1.74 &  13.47 & 28.18 & \textbf{0.556} \\
MMAudioSep (pretrain w/frozen) & 540M & \textbf{103.03} & 1.76 & \textbf{1.63} & \textbf{14.99} & \textbf{30.35} & \textbf{0.488} \\
\bottomrule
\end{tabular}
}
\end{table*}
\vspace{-1mm}
\vspace{-2mm}
\section{Experiments}
\label{sec:experiments}
\vspace{-1mm}
\subsection{Experimental Setup}
\label{ssec:exp_setup}
\vspace{-1mm}
\textbf{Training Dataset.}
To train MMAudioSep, we utilized the same dataset employed for the pretrained MMAudio, which totals approximately 2,500 hours. This includes 400 hours for Video+Audio+Label and 2,100 hours for Audio+Text.
We train on VGGSound \cite{chen2020vggsound} as the video-audio-text dataset. We use the first 8s of each video for training.
We also use AudioCaps \cite{kim2019audiocaps}, Clotho \cite{drossos2020clotho}, and WavCaps \cite{mei2023wavcaps} as the audio-text datasets. For short audios ($<$16s), we truncate them to 8s for training, as in VGGSound. For longer audios, we take up to five non-overlapping crops of 8s each. This results in a total of 951K audio clip-text pairs.\\
\textbf{Evaluation Dataset.} 
The AudioSep evaluation dataset is publicly accessible. Since our method also utilizes video queries, our evaluation focuses on samples for which the corresponding video was available on YouTube. For the purpose of separation evaluation, we then utilize 5,004 samples from the VGGSound-Clean dataset \cite{liu2023separate} and 1,000 samples from the MUSIC dataset \cite{Zhao_2018_ECCV}, which were both taken from the AudioSep website \footnote{\url{https://github.com/Audio-AGI/AudioSep}}. 
Each file contains mixture and target sound. In the VGGSound-Clean dataset, target and noise signals were sampled between -35dB and -25dB LUFS (Loudness Units Full Scale) and mixed. The average signal-to-noise ratio (SNR) of the VGGSound-Clean dataset is around 0 dB. In the MUSIC dataset, target and noise signals are mixed with an SNR at 0 dB. To construct a video-queried sound separation dataset, we append the corresponding video component to each file. Since the audio signals are uniformly 10 seconds long, black frames were inserted into the video if it was originally less than 10 seconds to ensure temporal consistency between audio and video.  
For the video-to-audio generation evaluation, we use the VGGSound \cite{chen2020vggsound} test set (\textasciitilde 15K videos) and label. \\
\textbf{Training Setup.} During the training process, SNR is randomly sampled from a uniform distribution ranging from -15 to 15 dB. The mixture signal is generated in an on-the-fly manner. The interference signal is randomly selected from the training dataset. The remaining training configuration adhered to the original MMAudio settings. In our experimental study, we trained three distinct model configurations: the one depicted in Fig. \ref{fig:arch}, a model developed entirely from the scratch, and a model that was initialized with pretrained weights, ensuring all parameters were set to be learnable.\\ 
\textbf{Inference Setup.} For inference, we use standard MMAudio settings. Euler's method was used for numerical integration over 25 steps, with classifier-free guidance strength at 4.5. \\
\textbf{Metrics.} 
As MMAudioSep is a generative-based approach, conventional sample-level objective metrics for sound separation tasks, such as the source-to-distortion ratio (SDR), are suboptimal for assessing the proposed system. Instead, we follow the evaluation protocol of FlowSep \cite{yuan2024flowsep}, employing metrics suited for generative tasks. These include Fréchet Audio Distance (FAD) \cite{kilgour2018fad} for feature distribution similarity and CLAP-based scores \cite{laion2023clap}. Specifically, CLAPScore evaluates the semantic alignment between the output audio and the text query, while CLAPScore$_{\textbf{A}}$ measures the similarity to the ground-truth target audio. 
To further evaluate generation quality, we use the av-benchmark \footnote{\url{https://github.com/hkchengrex/av-benchmark}}, assessing four key aspects: 
\begin{itemize}[leftmargin=9pt,itemsep=0pt, parsep=0pt, topsep=0pt]
\item Distribution Matching: The Fréchet Distance (FD) and Kullback-Leibler divergence (KL) are computed on the basis of PaSST \cite{koutini2021patchout}. The VGGish \cite{hershey2017vgg} is also used for FD (denoted as "FAD"). Note that we exclude PANNs \cite{kong2020panns} from FD and KL metric computation, as it has been reported as not being robust in some scenarios \footnote{\url{https://github.com/haoheliu/audioldm_eval}}.
\item Audio Quality: The Inception Score (IS) \cite{tim2016gan}, calculated with a PANNs classifier, is used to evaluate the quality and diversity of the generated audio.
\item AV Semantic Alignment: The alignment between the input video and generated audio is measured by the average cosine similarity of ImageBind \cite{girdhar2023imagebind} features (IB-Score).
\item AV Temporal Alignment: The synchronization between the input video and generated audio is measured by using Synchformer \cite{iashin2024synchformer} features to predict temporal misalignment (DeSync).
\end{itemize}
IS, IB-Score, and DeSync are also used to assess the quality and alignment in separation tasks.
All separation tasks were evaluated using 10-second audio segments, while generation tasks used 8-second segments.
\vspace{-3mm}
\subsection{Acquiring Sound Separation Functionality}
\label{ssec:main_exp}
\vspace{-1mm}
To demonstrate that the proposed method allows for conducting visual/text-queried sound separation based on a pretrained generation model, we evaluate the separation performance of MMAudioSep with different combinations of input queries. 
We compared MMAudioSep results from different query settings with AudioSep, which was trained on a 14,000-hour dataset supporting a 32kHz sampling rate, and FlowSep, which was trained on 1,680-hour dataset supporting a 16kHz sampling rate. We exclude CLIPSep and OmniSep as baselines. Training our larger model on their smaller datasets would likely cause overfitting, which would lead to an unfair comparison of the model architectures. 
We evaluated a 10-second duration to compare with AudioSep and FlowSep, utilizing their official released models and codes with default settings. Note that FlowSep's model differed from those trained on 1,680-hour datasets \footnote{\url{https://github.com/Audio-AGI/FlowSep}}. For the VGGSound-Clean dataset, we used class label as text query, while for the MUSIC dataset, we used Instruments label (e.g., accordion, acoustic guitar) as text query.

Table. \ref{tab:res_main} presents the sound separation benchmark results.
The findings suggest that MMAudioSep outperforms AudioSep, even when solely the text condition is used. 
Including both text and video conditions leads to performance being enhanced across nearly all metrics assessed. 
This highlights the strength of the multimodal training framework and the effectiveness of our generative-based model approach. It is important to highlight that the performance of FlowSep was significantly inferior to our method, particularly when evaluated on the MUSIC dataset. Although FlowSep is also a generative model, our approach is enhanced by a more extensive pretrained foundation and the incorporation of multimodal (video and text) queries. Additionally, as highlighted in their official repository, the FlowSep model available to the public is not identical to the version discussed in their paper, which could also explain the noted differences in performance.

Fig. \ref{fig:ex_ss_result} presents a visualization of the spectrograms for the mixture, ground-truth, and separated audio compared to AudioSep, where the frequency axis ranges up to 16kHz. Our model shows superior separation quality, particularly in its capacity to maintain detailed spectral features and reduce artifacts in the separated audio.
\vspace{-5mm}
\subsection{Preserving Video-to-Audio Generation Performance}
\label{ssec:v2a}
\vspace{-1mm}
To further investigate the behavior of MMAudioSep, we conducted an experiment to assess its performance on video-to-audio (V2A) generation. In this experiment, we injected random noise unrelated to the target audio as the input mixture, and we found that the output sound corresponded to the input video, demonstrating the original V2A functionality. 

Table. \ref{tab:res_vgg_all} shows the V2A benchmark results on the VGGSound test set.
Our approach utilized video and label information from VGGSound, along with white noise as a mixture signal.
Despite being trained for source separation, the model surpassed certain baseline generative models in specific metrics. 
This finding shows separation-based models can potentially be effective generative systems in multimodal contexts. 
Compared with MMAudio as a baseline model, MMAudioSep performed worse; however, it achieved results comparable to those of traditional V2A methods. 

Utilizing a pretrained model yielded higher performance than training from scratch by leveraging its learned multimodal information. Performance was further enhanced by freezing certain parameters, which retained visual and textual information more effectively than fine-tuning all parameters.

In our assessment of the audio outputs generated by the model, we detected occurrences of white noise. Although some sounds still resemble noise, the result in table \ref{tab:res_vgg_all} indicates that the model effectively operates as a V2A model. Future research will focus on developing methodologies to ensure stable and reliable sound generation.

\vspace{-2mm}
\section{Conclusion}
\label{sec:conclusion}
\vspace{-1mm}

This paper introduces MMAudioSep, an innovative approach to video/text-queried sound separation that leverages a pretrained video-to-audio generation model. The MMAudioSep model advances in multimodal sound separation by utilizing a fine-tuned pretrained MMAudio model with channel-concatenation conditioning. 
Its primary contributions include performing competitively with or superiorly to state-of-the-art models, maintaining video-to-audio generation capabilities, and enabling flexible multimodal querying. 
Furthermore, the introduction of generative model metrics for evaluating separation quality enhances its significance in the field. Our analysis indicates that MMAudioSep preserves the core functionalities of the original MMAudio, which include both sound separation and the capability for video-to-audio generation.
Future research will aim to enhance MMAudioSep towards achieving universal sound separation, thereby enabling the model to process a broader range of sound categories. We aim to investigate the potential of MMAudio as a foundational model for downstream applications, including sound event detection and visual-acoustic matching, by utilizing its multimodal representation capabilities.


\vfill\pagebreak

\clearpage
\bibliographystyle{IEEEbib}
{
\footnotesize
\bibliography{refs}
}
\vfill\pagebreak

\end{document}